\documentstyle[12pt]{article}
\newcommand{\be}{\begin{equation}}
\newcommand{\ee}{\end{equation}}
\newcommand{\bea}{\begin{array}}
\newcommand{\eea}{\end{array}}

\title{FURTHER REMARKS ON QUANTUM MECHANICS 
AND INTEGRABLE SYSTEMS}
\author{Robert Carroll\\
University of Illinois\\
Urbana, IL 61801\thanks{email:  rcarroll@symcom.math.uiuc.edu}}

\date{October, 1996}

\begin{document}
\bibliographystyle{plain}
\maketitle

\begin{abstract} 
Further formulas are presented involving quantum mechanics,
thermodynamics, and integrable systems.  Modifications of dispersionless
theory are developed.  
\end{abstract}


\section{INTRODUCTION}
\renewcommand{\theequation}{1.\arabic{equation}}\setcounter{equation}{0}

This paper is meant to be a sequel to \cite{ca} which in turn was based
on the fundamental paper \cite{fa}.  We will expand on some of the 
development in \cite{ca} and introduce a number of other heuristic
formulas.

\section{BACKGROUND}
\renewcommand{\theequation}{2.\arabic{equation}}\setcounter{equation}{0}

The point of departure is the Schr\"odinger equation
\be
{\cal H}\psi_E=-\frac{\hbar^2}{2m}\psi''_E+V(X)\psi_E=E\psi_E
\label{A}
\ee
where $X$ is the quantum mechanical (QM) space variable with $\psi'_E=
\partial\psi_E/\partial X$ and we write $\epsilon=\hbar/\sqrt{2m}\,\,
(E$ is assumed real).  In \cite{ca} we discussed the possible origin of
this from a Kadomtsev-Petviashvili (KP) situation $L^2_{+}\psi=\partial\psi/
\partial t_2$ where $L^2_{+}=\partial_x^2-v(x,t_i)$ and e.g. $\tau_2=
-i\sqrt{2m}T_2$ so $\partial_{t_2}=\epsilon\partial_{T_2}=-i\hbar
\partial_{\tau_2}$ (one writes $X=\epsilon x$ and 
$T_i=\epsilon t_i$ in the dispersionless theory).  This leads to an
approximation
\be
\epsilon^2\psi''_E-V(X,T_i)\psi_E\sim\epsilon\frac{\partial\psi_E}{\partial T_2}
=-i\hbar\frac{\psi_E}{\partial\tau_2} 
\label{B}
\ee
corresponding to the Schr\"odinger equation.  For the approximation one
assumes e.g. $v=v(x,t_i)\to v(\frac{X}{\epsilon},\frac{T_i}{\epsilon})
=V(X,T_i)
+O(\epsilon)$ (standard in dispersionless KP = dKP and certainly realizable
by quotients of homogeneous polynomials for example).  Further, when
$\psi_E=exp(S/\epsilon)$ for example, one has $\epsilon\psi'_E=S_X\psi_E$
with $\epsilon^2\psi''_E=\epsilon S_{XX}\psi_E+(S_X)^2\psi_E$ so in (\ref{B})
we are neglecting an $O(\epsilon)\psi_E$ term from $v$, and for $\psi_E
=exp(S/\epsilon)$ another $\epsilon S_{XX}\psi_E$ term is normally removed
in dispersionless theory.  Then for ${\cal H}$ independent of $\tau_2$ for
example one could assume $V$ is independent of $T_2$ and write formally
in (\ref{B}), $\hat{\psi}_E=exp(E\tau_2/i\hbar)\cdot\psi_E$, 
with ${\cal H}\psi_E
=E\psi_E$, which is (\ref{A}).  Since in the QM problem one does not
however run $\hbar\to 0$ (hence $\epsilon\not\to 0$) one could argue\
that these $O(\epsilon)$ terms should be retained, at least in certain
situations, and we will keep this in mind.  In particular one could ask
for $v(\frac{X}{\epsilon},\frac{T_i}{\epsilon})=V(X,T_i)+\epsilon\hat{V}
(X,T_i)+O(\epsilon^2)$ and retain the 
$\epsilon\hat{V}$ term along with $\epsilon
S_{XX}$, in requiring e.g. $S_{XX}=\hat{V}$.
\\[3mm]\indent
We list first a few of the equations from \cite{fa}, as written in 
\cite{ca}, without a discussion
of philosophy (some of which will
be mentioned later).  Thus ${\cal F}$ is a prepotential
and $\psi_E,\,\,\bar{\psi}_E=\psi_E^D$ satisfy (\ref{A}) with $\psi_E^D=
\partial{\cal F}/\partial\psi_E$.  
The Wronskian in (\ref{A}) is taken to be
$W=\psi'\bar{\psi}-\psi\bar{\psi}'=2\sqrt{2m}/i\hbar=2/i\epsilon$ and one
has ($\psi=\psi(X)$ and $X=X(\psi)$ with $X_{\psi}=\partial X/\partial\psi=
1/\psi'$)
\be
{\cal F}'=\psi'\bar{\psi};\,\,{\cal F}=\frac{1}{2}\psi\bar{\psi}+\frac
{X}{i\epsilon};\,\,
\frac{\partial\bar{\psi}}{\partial\psi}=\frac{1}{\psi}\left[\bar{\psi}-
\frac{2}{i\epsilon}X_{\psi}\right]
\label{FF}
\ee
($\psi$ always means $\psi_E$ but we omit the subscript occasionally
for brevity).
Setting $\phi=\partial{\cal F}/\partial(\psi^2)=\bar{\psi}/2\psi$ with
$\partial_{\psi}=2\psi\partial/\partial(\psi^2)$ and evidently
$\partial\phi/\partial
\psi=-(\bar{\psi}/2\psi^2)+(1/2\psi)(\partial\bar{\psi}/\partial\psi)$
one has a Legendre transform pair
\be
-\frac{X}{i\epsilon}=\psi^2\frac{\partial{\cal F}}{\partial(\psi^2)}
-{\cal F};\,\,-{\cal F}=\phi\frac{1}{i\epsilon}X_{\phi}-\frac{X}{i\epsilon}
\label{HH}
\ee
One obtains also
\be
|\psi|^2=2{\cal F}-\frac{2X}{i\epsilon}\,\,({\cal F}_{\psi}=\bar{\psi});\,\,
-\frac{1}{i\epsilon}X_{\phi}=\psi^2;\,\,{\cal F}_{\psi\psi}=\frac
{\partial\bar{\psi}}{\partial\psi}
\label{II}
\ee
Further from $X_{\psi}\psi'=1$ one has $X_{\psi\psi}\psi'+X^2_{\psi}\psi''
=0$ which implies
\be
X_{\psi\psi}=-\frac{\psi''}{(\psi')^3}=\frac{1}{\epsilon^2}\frac
{(E-V)\psi}{(\psi')^3}
\label{KK}
\ee
\be
{\cal F}_{\psi\psi\psi}=\frac{E-V}{4}({\cal F}_{\psi}-\psi\partial^2_{\psi}
{\cal F})^3=\frac{E-V}{4}\left(\frac{2X_{\psi}}{i\epsilon}\right)^3
\label{LL}
\ee
Although a direct comparison of (\ref{LL}) to the Gelfand-Dickey resolvant
equation ((\ref{AJJJ}) below) is not evident ($V'$ is lacking) a result of
T. Montroy which expands ${\cal F}_{\psi\psi\psi}$
shows that in fact (\ref{LL}) corresponds exactly to
\be
\epsilon^2{\cal F}'''+4(E-V)\left({\cal F}'-\frac{1}{i\epsilon}\right)
-2V'\left({\cal F}-\frac{X}{i\epsilon}\right)=0
\label{ZZZ}
\ee
which is (\ref{AJJJ}) since $\Xi=|\psi|^2=2{\cal F}-(2X/i\epsilon)$.
\\[3mm]\indent
Next there is a so-called eikonal transformation (cf. \cite{mf}) which can
be related to \cite{fa} as in \cite{ca}.  We consider
real $A$ and $S$ with
\be
\psi=Ae^{(i/\hbar)S};\,\,p=ASin(\frac{1}{\hbar}S);\,\,q=ACos(\frac{1}{\hbar}
S)
\label{OO}
\ee
Introduce new variables 
\be
\chi=A^2=|\psi|^2;\,\,\xi=\frac{1}{2\hbar}S
\label{PP}
\ee
and set ($'\sim\partial/\partial X$)
\be
K_1=\frac{1}{2}\int dX\left[\frac{\hbar^2}{2m}
\left(\frac{(\chi')^2}{4\chi}+4\chi
(\xi')^2\right)+V\chi\right]
\label{QQ}
\ee
Then it follows that
\be
\dot{\xi}=\frac{\hbar}{4m}\frac{(\sqrt{\chi})''}{\sqrt{\chi}}-\frac{\hbar}{m}
(\xi')^2-\frac{1}{2\hbar}V;\,\,\dot{\chi}=-\frac{2\hbar}{m}(\chi\xi')';\,\,
\delta p\wedge \delta q=\delta\xi\wedge\delta\chi=\tilde{\omega}
\label{RR}
\ee
Thus formally one has a Hamiltonian format with symplectic form as in
(\ref{RR}).
It is interesting to write down the connection between the $(S,A)$ or
$(\chi,\xi)$ type
variables and the variables from \cite{fa}.  
Take now $\psi=Aexp(iS/\epsilon)\,\,(\epsilon=\hbar/\sqrt{2m})$ with
$\xi\sim S/2\epsilon$; then
\be
{\cal F}=\frac{1}{2}\chi+\frac{X}{i\epsilon};\,\,{\cal F}'=
\psi'\bar{\psi} =\frac{1}{2}\chi'
+\frac{i}{\epsilon}P\chi
\label{TT}
\ee
for $S'=S_X=P$ and there is an interesting relation
\be
P\chi=-1\Rightarrow \delta\chi=-\frac{\chi}{P}\delta P
\label{UU}
\ee
Further from $\phi=(1/2)exp[-(2i/\epsilon)S]$ and $\psi^2=\chi exp(4i\xi)$ 
we have
\be
\psi^2\phi=\frac{1}{2}\chi=-\frac{1}{\epsilon}\phi X_{\phi};\,\,\xi=\frac
{S}{2\epsilon}=\frac{i}{4}log(2\phi)
\label{VV}
\ee
Now the theory of the Seiberg-Witten (SW) differential $\lambda_{SW}$ 
following \cite{bh,ca,ch,dd,ib,ke,ma,sd} for example 
involves finding a differential
$\lambda_{SW}$ of the form $QdE$ or $td\omega_0$ (in the spirit of
\cite{ke} or \cite{dd,ib} respectively) such that $d\lambda_{SW}=\omega$
is a symplectic form (cf. \cite{ch,ib} for some discussion).  In the
present context one can ask now whether the form $\tilde{\omega}$ of 
(\ref{RR}) makes any sense in such a context.  Evidently this is jumping
the gun since there is no Riemann surface in sight (see \cite{ca}
for a Riemann surface); the motivation to consider 
the matter here comes from the
following formulas which express $\tilde{\omega}$ nicely in terms of
the duality variables of \cite{fa}.
Thus a priori $\psi=\Re\psi+i\Im\psi$ has two components which are also
visible in $\psi=Aexp(iS/\epsilon)$ as $A$ and $S$.  The relation $P\chi
=\chi(\partial S/\partial X)=-1$ indicates a dependence between $A$
and $S'$ (but not $A$ and $S$) which is a consequence of the duality between
$\psi$ and $X$.  Then $2AS'\delta A+A^2\delta S'=0$ or $\delta S'
=-(2S'/A)\delta A\equiv (\delta S'/S')=-2(\delta A/A)$, whereas $\delta
\psi/\psi\sim 2(\delta A/A) +(i/\epsilon)\delta S$.  It follows that
$\Re(\delta\psi/\psi)=-(\delta S'/S')$ and $\Im(\delta \psi/\psi)=(\delta
S/\epsilon)$.  The sensible thing seems to be to look at the complex
dependence of $X(\psi)$ and $\psi(X)$ in terms of two real variables
and $\delta\xi\wedge\delta\chi$ will have a nice
form in transforming to the variables of \cite{fa}.  In particular
from $\psi^2\phi=(1/2)\chi$ with $\delta\chi=4\phi\psi\delta\psi+
2\psi^2\delta\phi$ we obtain
$(\delta\psi/\psi)=2(\delta\chi/\chi)-
(\delta\phi/\phi)$.
Hence one can write 
\be
\delta\xi\wedge\delta\chi=\frac{i}{4}\frac{\delta\phi}{\phi}
\wedge\chi\frac{\delta\chi}{\chi}=\frac{i}{2}\delta\phi\wedge\delta\psi^2
\sim\frac{i}{2}\delta\bar{\psi}\wedge\delta\psi
\label{XX}
\ee
(note $\delta\phi=(1/2\phi)\delta\bar{\psi}-(\bar{\psi}/2\psi^2)\delta\psi$)
and in an exploratory spirit the differentials $\lambda=(i/2)\phi\delta
\psi^2$ or $\lambda=(i/2)\psi^2\delta\phi$, along
with $\lambda=(i/2)\bar{\psi}\delta\psi$
or $\lambda=(i/2)\psi\delta\bar{\psi}$, might merit further consideration.
\\[3mm]\indent
We refer now to \cite{ci,cj,cm,tb} for dispersionless KP ($=$ dKP) and consider
here $\psi=exp[(1/\epsilon)S(X,T,\lambda)]$ instead of $\psi=Aexp(S/\epsilon)$
(more details are given in Section 3).
Thus $P=S'=S_X$ and $P^2=V-E$ 
but $E\not= \pm\lambda^2$ (unless otherwise stated) and
this does not define $S$ via $P=S_X$ unless we have a KdV situation
(which is not a priori desirable but will be used later with modifications);  
thus generally $\lambda$ is the $\lambda$ of $S(T_n,\lambda)$ from KP theory
and we recall that $\psi$ always means $\psi_E$ as in \cite{fa}.
One computes
easily
(recall $X_{\psi}=1/\psi'$ and $\psi'=(P/\epsilon)\psi$)
\be
\phi=\frac{1}{2}e^{-(2i/\epsilon)\Im S};\,\,\frac{1}{\epsilon}X_{\phi}=
-ie^{(2/\epsilon)S};\,\,X_{\psi}=\frac{\epsilon}{P}e^{-S/\epsilon}
\label{AAQ}
\ee
\be
\frac{1}{\epsilon}X_{\psi\psi}=\frac{E-V}{P^3}e^{-S/\epsilon};\,\,
{\cal F}_{\psi}=\bar{\psi}=e^{\bar{S}/\epsilon};\,\,
{\cal F}_{\psi\psi}=e^{-(2i/\epsilon)\Im S}-\frac{2}{iP}e^{-2S/\epsilon}
\label{ABQ}
\ee
Next from ${\cal F}'=\psi'\bar{\psi}=(P/\epsilon)exp[(2/\epsilon)\Re S]$
and $W=(2/i\epsilon)=(\psi'\bar{\psi}-\psi\bar{\psi}')=({\cal F}'
-\bar{{\cal F}}')$ we have $\Im{\cal F}'=-(1/\epsilon)$ and
from ${\cal F}=(1/2)\psi\bar{\psi}+(X/i\epsilon)$ we see that $\Im {\cal F}
=-(X/\epsilon)$.  In addition
\be
|\psi|^2=e^{(2/\epsilon)\Re S};\,\,\frac{S}{\epsilon}=\frac{1}{2}log|\psi|^2
-\frac{1}{2}log(2\phi)
\label{ACQ}
\ee
Finally 
\be
\bar{P}=\bar{S}_X=\epsilon\frac{\partial_X\bar{\psi}}{\bar{\psi}}=
\epsilon\frac{\partial\bar{\psi}}{\partial\psi}
\frac{\psi_X}{\bar{\psi}}=P\frac{\psi}{\bar{\psi}}\left[
\frac{\bar{\psi}}{\psi}-\frac{2}{iP\psi^2}\right]=P-\frac
{2}{i\psi\bar{\psi}}
\label{ADQ}
\ee
Summarizing one has
\be
\Im{\cal F}=-\frac{X}{\epsilon};\,\,\Re{\cal F}=\frac{1}{2}|\psi|^2=
\frac{1}{2}e^{\frac{2}{\epsilon}\Re S}=-\frac{1}{2\Im P}
\label{AGQ}
\ee
In the present situation $|\psi|^2=exp[(2/\epsilon)\Re S]$ and $2\phi=
exp[-(2i/\epsilon)\Im S]$ can play the roles of independent variables
(cf. (\ref{ACQ}).  The version here of $P\chi=-1$ is $\chi\Im P=-1$,
while $\psi^2\phi=(1/2)|\psi|^2=(1/2)\chi$ again, and
we obtain as above the formula (\ref{XX}).  Let us note also
from (\ref{ADQ}) that 
\be
\partial_{\lambda}{\cal F}=\frac{2}{\epsilon}\left({\cal F}-\frac{X}{i\epsilon}
\right)\Re S_{\lambda}\Rightarrow\partial_{\lambda}log\left({\cal F}-\frac
{X}{i\epsilon}\right)=\frac{2}{\epsilon}\Re{\cal M}
\label{AIQ}
\ee
where ${\cal M}$ is the dispersionless Orlov-Schulman operator (cf. \cite
{cg,ci,tb})
and $\lambda$ here is the $\lambda$ of KP theory.
Still another way to relate ${\cal F}$ and $F$ follows from the Gelfand-
Dickey resolvant equation (cf. \cite{ce}) for $\Xi=\psi\bar{\psi}$, namely,
in QM form
\be
\epsilon^2\Xi'''-4V\Xi'-2V'\Xi+4E\Xi'=0
\label{AJJJ}
\ee
(direct calculation).  Then note that for $L=\partial+\sum_1^{\infty}u_i
\partial^{-i},\,\,L^2_{+}=\partial^2+2u_1$, and $u_1=\partial^2log(\tau)$.
This implies $v=-2\partial^2log(\tau)$ here, from which $V=-2F_{XX}$ for
$\tau=exp[(1/\epsilon^2)F]$.  Now put this in (\ref{AJJJ}), using $\Xi=
2{\cal F}-(2X/i\epsilon),\,\,\Xi'=2{\cal F}'-(2/i\epsilon),\,\,\Xi''=
2{\cal F}''$, and $\Xi'''=2{\cal F}'''$, to obtain (cf. also (\ref{ZZZ}))
\be
\epsilon^2{\cal F}'''+\left({\cal F}'-\frac{1}{i\epsilon}\right)(8F''
+4E)+4F'''\left({\cal F}-\frac{X}{i\epsilon}\right)=0
\label{AKQ}
\ee
We will see in Section 3 how to embellish all this with a modification of
the dKP and dKdV theory.

\section{DISPERSIONLESS THEORY}
\renewcommand{\theequation}{3.\arabic{equation}}\setcounter{equation}{0}

\subsection{Classical framework for KP}

We give next a brief sketch of some ideas regarding dispersionless KP
(dKP) following mainly \cite{ci,cj,cm,kf,tb} to which we refer for 
philosophy.  We will make various notational adjustments as we go along 
and subsequently will modify some of the theory.  One
can think of fast and slow variables with $\epsilon x=X$ and $\epsilon t_n=
T_n$ so that $\partial_n\to\epsilon\partial/\partial T_n$ and $u(x,t_n)
\to\tilde{u}(X,T_n)$ to obtain from the KP equation $(1/4)u_{xxx}+3uu_x
+(3/4)\partial^{-1}\partial^2_2u=0$ the equation $\partial_T\tilde{u}=3
\tilde{u}\partial_X\tilde{u}+(3/4)\partial^{-1}(\partial^2\tilde{u}/
\partial T_2^2)$ when $\epsilon\to 0$ ($\partial^{-1}\to(1/\epsilon)
\partial^{-1}$).  In terms of hierarchies the theory can be built around the
pair $(L,M)$ in the spirit of \cite{cg,ci,tb}.  Thus writing $(t_n)$ for
$(x,t_n)$ (i.e. $x\sim t_1$ here) consider
\be
L_{\epsilon}=\epsilon\partial+\sum_1^{\infty} u_{n+1}(\epsilon,T)
(\epsilon\partial)^{-n};\,\,M_{\epsilon}=\sum_1^{\infty}nT_nL^{n-1}_{\epsilon}
+\sum_1^{\infty}v_{n+1}(\epsilon,T)L_{\epsilon}^{-n-1}
\label{YA}
\ee
Here $L$ is the Lax operator $L=\partial+\sum_1^{\infty}u_{n+1}\partial^{-n}$
and $M$ is the Orlov-Schulman operator defined via $\psi_{\lambda}=M\psi$.
Now one assumes $u_n(\epsilon,T)=U_n(T)+O(\epsilon)$, etc. and 
set (recall $L\psi=\lambda\psi$)
$$
\psi=\left[1+O\left(\frac{1}{\lambda}\right)\right]exp
\left(\sum_1^{\infty}\frac{T_n}
{\epsilon}\lambda^n\right)=exp\left(\frac{1}
{\epsilon}S(T,\lambda)+O(1)\right);$$
\be
\tau=exp\left(\frac{1}{\epsilon^2}F(T)+O\left(\frac{1}{\epsilon}\right)
\right)
\label{YB}
\ee
We recall that $\partial_nL=[B_n,L],\,\,B_n=L^n_{+},\,\,\partial_nM
=[B_n,M],\,\,[L,M]=1,\,\,L\psi=\lambda\psi,\,\,\partial_{\lambda}\psi
=M\psi,$ and $\psi=\tau(T-(1/n\lambda^n))exp[\sum_1^{\infty}T_n\lambda^n]/
\tau(T)$.  Putting in the $\epsilon$ and using $\partial_n$ for
$\partial/\partial T_n$ now, with $P=S_X$, one obtains
\be
\lambda=P+\sum_1^{\infty}U_{n+1}P^{-n};\,\,
P=\lambda-\sum_1^{\infty}P_i\lambda^{-1};
\label{YC}
\ee
$${\cal
M}=\sum_1^{\infty}nT_n\lambda^{n-1}+\sum_1^{\infty}V_{n+1}\lambda^{-n-1};
\,\,\partial_nS={\cal B}_n(P)\Rightarrow \partial_nP=\hat{\partial}
{\cal B}_n(P)$$
where $\hat{\partial}\sim \partial_X+(\partial P/\partial X)\partial_P$
and $M\to {\cal M}$.
Note that one assumes also $v_{i+1}(\epsilon,T)=V_{i+1}(T)+O(\epsilon)$; 
further
for $B_n=\sum_0^nb_{nm}\partial^m$ one has ${\cal B}_n=\sum_0^nb_{nm}
P^m$ (note also $B_n=L^n+\sum_1^{\infty}\sigma_j^nL^{-j}$).  
We list a few additional formulas which are 
easily obtained (cf. \cite{ci}); thus, writing $\{A,B\}=\partial_PA\partial A
-\partial A\partial_PB$ one has
\be
\partial_n\lambda=\{{\cal B}_n,\lambda\};\,\,\partial_n{\cal M}
=\{{\cal B}_n,{\cal M}\};\,\,\{\lambda,{\cal M}\}=1
\label{YD}
\ee
Now we can write $S=\sum_1^{\infty}T_n\lambda^n+\sum_1^{\infty}S_{j+1}
\lambda^{-j}$ with $S_{n+1}=-(\partial_nF/n),\,\,
\partial_mS_{j+1}=\tilde{\sigma}_j^m=-(F_{mn}/n),
\,\,V_{n+1}=-nS_{n+1}$,
and $\partial_{\lambda}S={\cal M}\,\, (
\sigma_j^m\to\tilde{\sigma}_
j^m$).  Further 
\be
{\cal B}_n=\lambda^n+\sum_1^{\infty}\partial_nS_{j+1}\lambda^{-j};\,\,
\partial S_{n+1}\sim -P_n\sim -\frac{\partial V_{n+1}}{n}\sim
-\frac{\partial\partial_n F}{n}
\label{YE}
\ee
\indent
We sketch next a few formulas from \cite{kf} (cf. also \cite{ci}).  First
it will be important to rescale the $T_n$ variables and write 
$t'=nt_n,\,\,T_n'=nT_n,\,\,
\partial_n=n\partial'_n=n(\partial/\partial T'_n)$.  Then
\be
\partial'_nS=\frac{\lambda^n_{+}}{n};\,\,\partial'_n\lambda=\{{\cal Q}_n,
\lambda\}\,\,({\cal Q}_n=\frac{{\cal B}_n}{n});
\label{YF}
\ee
$$\partial'_nP=\hat{\partial}{\cal Q}_n=\partial{\cal Q}_n+\partial_P
{\cal Q}_n\partial P;\,\,\partial'_n{\cal Q}_m-\partial'_m{\cal Q}_n=
\{{\cal Q}_n,{\cal Q}_m\}$$
Now think of $(P,X,T'_n),\,\,n\geq 2,$ as basic Hamiltonian variables
with $P=P(X,T'_n)$.  Then $-{\cal Q}_n(P,X,T'_n)$ will serve as a
Hamiltonian via
\be
\dot{P}'_n=\frac{dP'}{dT'_n}=\partial{\cal Q}_n;\,\,\dot{X}'_n=\frac
{dX}{dT'_n}=-\partial_P{\cal Q}_n
\label{YG}
\ee
(recall the classical theory for variables $(q,p)$ involves $\dot{q}=
\partial H/\partial p$ and $\dot{p}=-\partial H/\partial q$).  The function
$S(\lambda,X,T_n)$ plays the role of part of a generating function 
$\hat{S}$ for the Hamilton-Jacobi theory with action angle variables 
$(\lambda,-\xi)$ where
\be
PdX+{\cal Q}_ndT'_n=-\xi d\lambda-K_ndT'_n+d\hat{S};\,\,K_n=-R_n=-
\frac{\lambda^n}{n};
\label{YH}
\ee
$$\frac{d\lambda}{dT'_n}=\dot{\lambda}'_n=\partial_{\xi}R_n=0;\,\,\frac
{d\xi}{dT'_n}=\dot{\xi}'_n=-\partial_{\lambda}R_n=-\lambda^{n-1}$$
(note that $\dot{\lambda}'_n=0\sim\partial'_n\lambda=\{{\cal Q}_n,
\lambda\}$).  To see how all this fits together we write
\be
\frac{dP}{dT'_n}=\partial'_nP+\frac{\partial P}{\partial X}\frac{dX}
{dT'_n}=\hat{\partial}{\cal Q}_n+\frac{\partial P}{\partial X}\dot
{X_n}'=\partial{\cal Q}_n+\partial P\partial_P{\cal Q}_n+\partial P
\dot{X}'_n
\label{YI}
\ee
This is compatible with (\ref{YG}) and Hamiltonians $-{\cal Q}_n$.  Furthermore
one wants
\be
\hat{S}_{\lambda}=\xi;\,\,\hat{S}_X=P;\,\,\partial'_n\hat{S}=
{\cal Q}_n-R_n
\label{YJ}
\ee
and from (\ref{YH}) one has
\be
PdX+{\cal Q}_ndT'_n=-\xi d\lambda+R_ndT'_n+\hat{S}_XdX+\hat{S}_
{\lambda}d\lambda+\partial'_n\hat{S}dT'_n
\label{YK}
\ee
which checks.  We note that $\partial'_nS={\cal Q}_n={\cal B}_n/n$ and
$S_X=P$ by constructions and definitions.  Consider $\hat{S}=S-\sum_2^{\infty}
\lambda^nT'_n/n$.  Then $\hat{S}_X=S_X=P$ and $\hat{S}_n'=S_n'-R_n=
{\cal Q}_n-R_n$ as desired with $\xi=\hat{S}_{\lambda}=S_{\lambda}-
\sum_2^{\infty}T'_n\lambda^{n-1}$.  It follows that
$\xi\sim{\cal M}-\sum_2^{\infty}T'_n\lambda^{n-1}=X+\sum_1^{\infty}V_{i+1}
\lambda^{-i-1}$.  If $W$ is the gauge operator such that $L=W\partial W^{-1}$
one sees easily that
\be
M\psi
=W\left(\sum_1^{\infty}kx_k\partial^{k-1}\right)W^{-1}\psi=\left(G+
\sum_2^{\infty}kx_k\lambda^{k-1}\right)\psi
\label{YL}
\ee
from which follows that $G=WxW^{-1}\to\xi$.  This shows that $G$ is
a very fundamental object and this is encountered in various places
in the general theory (cf. \cite{cg,ci}).

\subsection{Dispersonless theory for KdV}

Following \cite{cb,ce,ci} we write
\be
L^2=L^2_{+}=\partial^2+q=\partial^2-u\,\,(q=-u=2u_2);\,\,q_t-6qq_x-q_{xxx}=0;
\label{YM}
\ee
$$B=4\partial^3+6q\partial+3q_x;\,\,L^2_t=[B,L^2];\,\,q=-v^2-v_x\sim
v_t+6v^2v_x+v_{xxx}=0$$
($v$ satisfies the mKdV equation).  KdkV is Galilean invariant ($x'=x-
6\lambda t,\,\,t'=t,\,\,u'=u+\lambda$) and consequently one can consider
$L+\partial^2+q-\lambda=(\partial+v)(\partial-v,\,\,q-\lambda=-v_x-v^2,\,\,
v=\psi_x/\psi,$ and $-\psi_{xx}/\psi=q-\lambda$ or $\psi_{xx}+q\psi=\lambda\psi$
(with $u'=u+\lambda\sim q'=q-\lambda$).  The $v$ equation in (\ref{YM}) becomes
then $v_t=\partial(-6\lambda v+2v^3-v_{xx})$ and for $\lambda=-k^2$ one
expands for $\Im k>0,\,\,|k|\to\infty$ to get
$(\clubsuit\clubsuit\clubsuit)\,\,
v\sim ik+\sum_1^{\infty}(v_n/(ik)^n)$.  The $v_n$ are conserved
densities and with $2-\lambda=-v_x-v^2$ one obtains
\be
p=-2v_1;\,\,2v_{n+1}=-\sum_1^{n-1}v_{n-m}v_m-v'_n;\,\,2v_2=-v'_1
\label{YN}
\ee
Next for $\psi''-u\psi=-k^2\psi$
write $\psi_{\pm}\sim exp(\pm ikx)$ as $x\to\pm\infty$.  Recall also the 
transmission and reflection coefficient formulas (cf. \cite{ce})
$T(k)\psi_{-}=R(k)\psi_{+}+\psi_{+}(-k,x)$ and $T\psi_{+}=R_L\psi_{-}+
\psi_{-}(-k,x)$.  Writing e.g. $\psi_{+}=exp(ikx+\phi(k,x))$ with
$\phi(k,\infty)=0$ one has $\phi''+2ik\phi'+(\phi')^2=u$.  Then
$\psi'_{+}/\psi_{+}=ik+\phi'=v$ with $q-\lambda=-v_x-v^2$.  Take then
\be
\phi'=\sum_1^{\infty}\frac{\phi_n}{(2ik)^n};\,\,v\sim ik+\phi'=ik+
\sum\frac{v_n}{(ik)^n}\Rightarrow\phi_n=2^nv_n
\label{YO}
\ee
Furthermore one knows (cf. \cite{ca})
\be
log T=-\sum_0^{\infty}\frac{c_{2n+1}}{k^{2n+1}};\,\,c_{2n+1}=\frac{1}{2\pi i}
\int_{-\infty}^{\infty}k^{2n}log(1-|R|^2)dk
\label{YP}
\ee
(assuming for convenience that there are no bound states).  Now for
$c_{22}=R_L/T$ and $c_{21}=1/T$ one has as $k\to -\infty\,\,(\Im k>0)$ the
behavior $\psi_{+}exp(-ikx)\to c_{22}exp(-2ikx)+c_{21}\to c_{21}$.  Hence
$exp(\phi)\to c_{21}$ as $x\to -\infty$ or $\phi(k,-\infty)=-log T$ which
implies
\be
\int_{-\infty}^{\infty}\phi'dx=log T=\sum_1^{\infty}\int_{-\infty}^{\infty}
\frac{\phi_ndx}{(2ik)^n}
\label{YQ}
\ee
Hence $\int\phi_{2m}dx=0$ and $c_{2m+1}=-\int\phi_{2m+1}dx/(2i)^{2m+1}$.
The $c_{2n+1}$ are related to Hamiltonians $H_{2n+1}=\alpha_nc_{2n+1}$
as in \cite{cb,cg} and thus the conserved densities $v_n\sim \phi_n$ give
rise to Hamiltonians $H_n$ (n odd).  There are action angle variables
$P=klog|T|$ and $Q=\gamma arg(R_L/T)$ with Poisson structure $\{F,G\}\sim
\int(\delta F/\delta u)\partial (\delta G/\delta u)dx$ (we omit the
second Poisson structure here).  
\\[3mm]\indent
Now look at the dispersionless theory based on $k$ where $\lambda^2
\sim(ik)^2=-k^2$.  One obtains for $P=S_X,\,\,P^2+q=-k^2$, and we write
${\cal P}=(1/2)P^2+p=(1/2)(ik)^2$ with $q\sim 2p\sim 2u_2$.  One has
$\partial k/\partial T_{2n}=\{(ik)^{2n},k\}=0$ and from $ik=P(1+qP^{-2})^
{1/2}$ we obtain
\be
ik=P\left(1+\sum_1^{\infty}{\frac{1}{2}\choose m}q^mP^{-2m}\right)
\label{YR}
\ee
(cf. (\ref{YC}) with $u_2=q/2$).  The flow equations become then
\be
\partial'_{2n+1}P=\hat{\partial}{\cal Q}_{2n+1};\,\,\partial'_{2n+1}(ik)
=\{{\cal Q}_{2n+1},ik\}
\label{YS}
\ee
Note here some rescaling is needed since we want $(\partial^2+q)^{3/2}_{+}=
\partial^3+(3/2)q\partial+(3/4)q_x=B_3$ instead of our previous $B_3\sim
4\partial^3 +6q\partial+3q_x$.  Thus we want ${\cal Q}_3=(1/3)P^3+(1/2)qP$
to fit the notation above.  The Gelfand-Dickey resolvant 
coefficients are defined via $R_s(u)=(1/2)Res(\partial^2-u)^{s-(1/2)}$
and in the dispersionless picture 
$R_s(u)\to (1/2)r_{s-1}(-u/2)$ (cf. \cite{ci}) where
$$
r_n=Res(-k^2)^{n+(1/2)}=\left(
\begin{array}{c}
n+(1/2)\\
n+1
\end{array}\right)q^{n+1}=
\frac{(n+1/2)\cdots(1/2}{(n+1)!}q^{n+1};$$
\be
\,\,2\partial_qr_n=(2n+1)r_{n-1}
\label{YT}
\ee
The inversion formula corresponding to (\ref{YC}) is $P=ik-\sum_1^{\infty}
P_j(ik)^{-j}$ and one can write
\be
\partial'_{2n+1}(P^2+q)=\partial'_{2n+1}(-k^2);\,\,\partial'_{2n+1}q
=\frac{2}{2n+1}\partial r_n=\frac{2}{2n+1}\partial_qr_nq_X=
q_Xr_{n-1}
\label{YU}
\ee
Note for example $r_0=q/2,\,\,r_1=3q^2/8,\,\,r_2=5q^3/16,\cdots$ and
$\partial'_Tq=q_Xr_0=(1/2)qq_X$ (scaling is needed in (\ref{YM}) here for
comparison).  Some further calculation gives for $P=ik-\sum_1^{\infty}
P_n(ik)^{-n}$ 
\be
P_n\sim-v_n\sim-\frac{\phi_n}{2^n};\,\,c_{2n+1}=(-1)^{n+1}\int_
{-\infty}^{\infty}P_{2n+1}(X)dX
\label{YV}
\ee
The development above actually gives a connection between inverse
scattering and the dKdV theory (cf. \cite{ci,cj,cm} for more on this).

\subsection{Another look at dKP}

The dKP theory as in \cite{ci,cj,kf,tb} involves a parameter $\epsilon\to
0$ and we recall $L=\partial+\sum_1^{\infty}u_{n+1}(t)\partial^{-n}\to
L_{\epsilon}=\epsilon\partial +\sum_1^{\infty}u_{n+1}(\epsilon,T)(\epsilon
\partial)^{-n}$ where $t\sim (t_k),\,\,T\sim (T_k),$ and $X=T_1$ with
$u_{n+1}(\epsilon,T)=U_{n+1}(T)+O(\epsilon)$ as in Section 2.  Then for
$\psi=exp(S/\epsilon)$ one has $L\psi=\lambda\psi\to\lambda= P+\sum_1^{\infty}
U_{n+1}P^{-n}$ where $P=S_X$ with $S=S(X,T_k,\lambda)\,\,(k\geq 2)$.
Here all the terms which are $O(\epsilon)$ are passed to zero and in
view of $\epsilon\not\to 0$ in the QM situation where $\epsilon=\hbar/\sqrt
{2m}$ one thinks of rewriting some of the dKP theory in order to retain
$O(\epsilon)$ terms at least (and dropping $O(\epsilon^2)$ terms).  
We will call this $dKP_{\epsilon}$ theory.  First
as indicated in Section 2 we could take e.g. $u_{n+1}(T/\epsilon)=
U_{n+1}(T)+\epsilon\hat{U}_{n+1}(T)+O(\epsilon^2)$.  Then for 
$\tilde{S}=S^0+\epsilon S^1$ with $\psi=exp(\tilde{S}/\epsilon)=exp
[(S^0/\epsilon)+S^1]$ we have $\tilde{P}=\partial\tilde{S}=S^0_X+
\epsilon S^1_X=P+\epsilon P^1$ so that $\epsilon\partial\psi=\tilde{P}
\psi=(P+\epsilon P^1)\psi,\,\,\epsilon^2\partial^2\psi=(\epsilon P_X
+\epsilon^2P_X^1)\psi +2\epsilon P^1P\psi+P^2\psi+\epsilon^2 (P^1)^2\psi$,
etc. 
along with $\epsilon\partial(\psi/\tilde{P})=
-\epsilon(\tilde{P}_X/\tilde{P}^2)\psi+\psi=\psi
-\epsilon((P_X/P^2)\psi+O(\epsilon^2)$
from which $(\epsilon\partial)\psi\to\psi$ or $(\epsilon\partial)^{-1}\psi
\to \psi/\tilde{P}$ in some sense.  Continuing such calculations we obtain
terms of $O(\epsilon)$ in $(1/\psi)(\epsilon\partial)^{-n}\psi$ of the form
\be
\frac{1}{\psi}
{n+1 \choose 2}(\epsilon\partial)^{-n}\left(\frac{P_X\psi}{P^2}\right)
\label{one}
\ee
Hence from $L\psi=\lambda\psi$ we get to first order
$$
\lambda=P+\sum_1^{\infty}U_{n+1}P^{-n}+\epsilon\left(P^1+\sum_1^{\infty}
\hat{U}_{n+1}P^{-n}\right.+$$
\be
+\left.\sum_1^{\infty}\left\{\frac{n(n+1)}{2\psi}\right\}
(\epsilon\partial)^{-n}\left[\frac{P_X\psi}{P^2}\right]-
P^1\sum_1^{\infty}nU_{n+1}P^{-n-1}\right)
\label{two}
\ee
\indent
As for ${\cal B}_n=\lambda^n_{+}$ we write $B_n^{\epsilon}\psi=\sum_0^n
b_{nm}^{\epsilon}(\epsilon\partial)^m\psi$ with $b^{\epsilon}_{nm}=
b_{nm}+\epsilon \hat{b}_{nm}$ and here we have for $\psi=
exp(\tilde{S}/\epsilon)$
a simple calculation of $\epsilon$ terms.  Writing ${\cal B}_n(P)=
\sum_0^nb_{nm}P^m$ we have, with a little calculation
\be
\frac{B_n^{\epsilon}\psi}{\psi}={\cal B}_n(P)+\epsilon 
P_X\left[\sum_2^nb_{nk}
\left(\frac{k(k-1)}{2}\right)P^{k-2}\right]+
\label{three}
\ee
$$+\epsilon \left\{P^1\sum_1^nkb_{nk}P^{k-1}+
\sum_0^n\hat{b}_{nm}P^k\right\}+
O(\epsilon^2)$$
and we write this as
$\tilde{{\cal B}}_n={\cal B}_n(P)+\epsilon {\cal B}_n^1(P,P^1)+O(\epsilon^2)$.
Both (\ref{two}) and (\ref{three}) have $P_X$ terms which seem
inappropriate in dealing with a HJ theory where $(X,P)$ are considered
independent and we will deal with this later.
\\[3mm]\indent
We recall now from Section 3.1 that in the traditional dKP theory
\be
S=\sum_1^{\infty}T_n\lambda^n+\sum_1^{\infty}S_{j+1}\lambda^{-j};\,\,
{\cal B}_n=\partial_nS=\lambda^n+\sum_1^{\infty}\partial_nS_{j+1}
\lambda^{-j}
\label{four}
\ee
and via $log\psi=(S/\epsilon)+O(1)\sim log\tau[\epsilon,T_n-(\epsilon/
n\lambda^n)]-log\tau+\sum_1^{\infty}T_n\lambda^n/\epsilon$ with $log\tau
=(F/\epsilon^2)+O(1/\epsilon)$ and $F[T_n-(\epsilon/n\lambda^n)]-F(T_n)\sim
-\epsilon\sum_1^{\infty}(\partial_nF/n\lambda^n)+O(\epsilon^2)$ one 
obtains $S_{n+1}=-(\partial_nF/n)$.  Consider now the next order terms
via $F$, i.e.
\be
F\left(T_n-\frac{\epsilon}{n\lambda^n}\right)-F(T_n)=-\epsilon\sum_1^{\infty}
\left(\frac{\partial_nF}{n\lambda^n}\right)+\frac{\epsilon^2}{2}\sum
\left(\frac{F_{mn}}{mn}\right)\lambda^{-m-n}+O(\epsilon^3)
\label{five}
\ee
Thus $\Delta log\tau=(1/\epsilon^2)\Delta F$ has $O(1)$ terms $(1/2)\sum
(F_{mn}/mn)\lambda^{-m-n}$ which correspond to the $O(1)$ terms in $log\psi$.
Hence we have a natural way of writing
$\tilde{S}=S^0+\epsilon S^1$ with $S^0=S$ as in (\ref{four}) and
\be
S^1=\frac{1}{2}\sum\left(\frac{F_{mn}}{nm}\right)\lambda^{-m-n}
\label{six}
\ee
In accord with $\partial_n\tilde{S}=
\tilde{{\cal B}}_n$ we should have now from (\ref{three})
\be
\partial_nS^1=P_X\left[\sum_2^nb_{nk}\left(\frac{k(k-1)}{2}\right)P^{k-2}
\right]+P^1\sum_2^nkb_{nk}P^{k-1}={\cal B}_n^1
\label{seven}
\ee
where indeed
\be
\partial_n\psi=B_n^{\epsilon}\psi\Rightarrow\partial_n
\tilde{S}\psi=(\partial_nS^0
+\epsilon\partial_nS^1)\psi=\tilde{{\cal B}}_n\psi=
{\cal B}_n\psi+\epsilon{\cal B}_n^1\psi
\label{eight}
\ee
In particular one can write now
\be
\tilde{S}_X\sim P+\frac{\epsilon}{2}\sum\left(\frac{F_{1mn}}{nm}\right)\lambda^
{-m-n}
\label{nine}
\ee
\indent
Consider next equations
(\ref{AAQ}) - (\ref{AKQ}).  Thus take now 
$\tilde{S}=S^0+\epsilon S^1$ with $S^0=S$
given by (\ref{four}) and $S^1$ by (\ref{six}).  Take $P=S_X^0$ and write
\be
\tilde{P}=\tilde{S}_X=P+\epsilon P^1;\,\,P^1=\partial_XS^1
\label{alpha}
\ee
so $P^1$ is given by (\ref{nine}).  
Take $\psi=exp[(1/\epsilon)\tilde{S}(X,T,\lambda)]$ now as in 
(\ref{AAQ}) and then the equations (\ref{AAQ}) - (\ref{AIQ}) hold with 
$P$ replaced by $\tilde{P}=\tilde{S}_X=
P+\epsilon P^1$ where $\tilde{S}=S^0+\epsilon S^1$.
For (\ref{AJJJ}) and (\ref{AKQ}) one should think of $V=-2F_{XX}$ in the form
$V=V^0+\epsilon \hat{V}$ with $S^0_{XX}+2S_XS^1_X=\hat{V}$ 
but the connection here to $F$ is not so clear
(actually in the $dKdV_{\epsilon}$ situation to be examined below this
will require $\hat{V}=0$ which will be a constraint - see (\ref{OT})).
In fact one could formally insert some ad hoc
$1/\epsilon$ terms in $F/\epsilon^2$ via
$F=F^0+\epsilon F^1$ in which case we modify (\ref{five}) - (\ref{six}) as
follows.  Set $\Delta F=\Delta F^0+\epsilon\Delta F^1$ so that in (\ref{five})
\be
\Delta F=-\epsilon\sum\left(\frac{\partial_nF^0}{n\lambda^n}\right)+
\epsilon^2\left[\frac{1}{2}\sum\left(\frac{F_{mn}^0}{mn}\right)\lambda^{-m-n}
-\sum\left(\frac{\partial_nF^1}{n}\right)\lambda^{-n}\right]+O(\epsilon^3)
\label{OB}
\ee
This gives, instead of (\ref{six}) and (\ref{nine}),
\be
S^1=\frac{1}{2}\sum\left(\frac{F^0_{mn}}{mn}\right)\lambda^{-m-n}-
\sum\left(\frac{\partial_nF^1}{n}\right)\lambda^{-n};
\label{OC}
\ee
$$\tilde{P}=P+\epsilon\left[\frac{1}{2}\sum\left(\frac{F^0_{1mn}}{mn}\right)
\lambda^{-m-n}-\sum\left(\frac{F^1_{1n}}{n}\right)\lambda^{-n}\right]$$
Further one has
\be
V=V^0+\epsilon\hat{V}=-2F_{XX}=-2F^0_{XX}-2\epsilon F^1_{XX}
\label{OD}
\ee
along with $\hat{V}=P_X+2PP^1=-2F^1_{XX}$, requiring some constraints
on $F^1$ which we do not specify here (in fact for the most interesting
situation of $dKdV_{\epsilon}$ with $\hat{V}=0$ one gets no milage
in this manner and $F^1=0$ is indicated).
Evidently one can now carry out such first order calculations for all 
quantities arising in dKP and we will return to this later.

\section{HAMILTON JACOBI THEORY}
\renewcommand{\theequation}{4.\arabic{equation}}\setcounter{equation}{0}

We will see that $dKdV_{\epsilon}$ has some attractive features whereas
dKdV is untenable.  Thus note first that
the equation ${\cal F}=(1/2)\psi\bar{\psi}+(X/i\epsilon)\sim (1/2)exp
[(2/\epsilon)\Re S] +(X/i\epsilon)$ has $\epsilon$ at various levels
which is confusing.  Moreover $|\psi|^2=exp[(2/\epsilon)\Re S]$ should
be bounded by $1$ which suggests a $dKP_{\epsilon}$ format
with $S\to \tilde{S}=S^0+\epsilon S^1\,\,(S^0\sim S),\,\, \Re S^0=0$, and
\be
|\psi|^2=e^{2\Re S^1}=exp\left[\Re\sum\left(\frac{F^0_{mn}}{mn}\right)
\lambda^{-m-n}\right]
\label{OF}
\ee
For this to occur we need 
\be
\Re S^0=\Re\sum_1^{\infty}T_n\lambda^n+\Re\sum_1^{\infty}S_{j+1}\lambda^
{-j}=0
\label{OG}
\ee
where one expects $S_{j+1}=-(\partial_jF^0/j)$ to be real.  This suggests
that it would be productive to think of KdV after all with $\lambda=ik$
imaginary, $T_{2n}=0,$ and $\partial_{2n}F^0=0$ as indicated below (so
$S_{2n+1}=0$ and only $\lambda^{-j}$ terms occur in (\ref{OG}) for $j$
odd).  Arguments against dKdV are indicated below but $dKdV_{\epsilon}$
seems appropriate as subsequently indicated.  
For $dKdV_{\epsilon}$  
one establishes $F^0_{m,2n}=0$ as in \cite{cj} (cf. below)
so in (\ref{OF}) one only has terms
\be
\frac{F^0_{(2m+1)(2n+1)}}{(2n+1)(2m+1)}\cdot\lambda^{-2(m+n)-2}
\label{OH}
\ee
which would be real for $\lambda=ik$.  Thus $S^0$ and $P=S^0_X$ are
imaginary while $S^1$ and $P^1=\partial_X S^1$ are real. 
In order to exhibit this context in a broader sense we digress
here to the Hamilton Jacobi (HJ) picture as in \cite{ci,kf}.

\subsection{Interaction with HJ theory}

Consider the Hamilton Jacobi (HJ) theory of Section 3.1
in conjunction
with the formulas of Section 2.  As background let us
assume we are considering a Schr\"odinger equation which in fact
arises from a KP equation (or KdV possibly) as indicated in Section 2.
Then one defines a prepotential ${\cal F}$ and it automatically must
have relations to a free energy as in (\ref{AKQ}) etc. The
HJ dynamics involve $T_n=nT'_n\,\,(\partial_n
=n\partial'_n)$ with
\be
\partial_nP=\partial{\cal B}_n+\partial_P{\cal B}_n\partial P;\,\,
\dot{P}_n=\frac{dP}{dT_n}=\partial{\cal B}_n;\,\,\dot{X}_n=\frac{dX}
{dT_n}=-\partial_P{\cal B}_n
\label{LK}
\ee
where ${\cal B}_n=\lambda^n_{+}=\sum_0^nb_{nj}P^j$
(cf. (\ref{YF}), (\ref{YG})) and this serves as another vehicle to put
$X$ in the picture so that comparison with ${\cal F}$ can be made. 
We emphasize here the strong nature
of the dependence $\psi=\psi(X)$ and $X=X(\psi)$ with all other quantities
dependent on $X$ or $\psi$ in \cite{fa}
(along with $X=X(T)$ arising in the HJ theory)
and this will introduce constraints.
The action term $S$ is given a priori as $S(X,T,\lambda)$ with $\lambda$
given via (\ref{YC}) as a function of $P$ and we recall that $\lambda$ and
$\xi=\hat{S}_{\lambda}=S_{\lambda}-\sum_2^{\infty}nT_n\lambda^{n-1}$ are
action-angle variables with $d\lambda/dT_n=0$ and $d\xi/dT_n=-n\lambda^{n-1}$.
For the moment we do not use $dKP_{\epsilon}$ 
or $dKdV_{\epsilon}$ and it will become apparent
why they are needed.
Note that the $b_{nj}=b_{nj}(U)$ should
be real and the conditions under which the formulas of 
\cite{fa} are valid with $E=\pm\lambda^2$ real involve $\lambda$ either
real or pure imaginary.  
A little thought shows that a KdV situation here with $\lambda
=ik,\,\,\lambda^2=-k^2=-E$ would seem to work
and we try this here to see what a KdV situation 
(first without $dKdV_{\epsilon}$)
will involve. We 
will have then $P$ purely imaginary with $U_j$ and $P_j$ real
and note that only odd powers of $P$ or 
$k$ appear in (\ref{YR}).  Look now at (\ref{YR}), i.e.
$ik=P(1+\sum_1^{\infty}U_mP^{-2m})$, and for $P=iQ$ we see
that $(ik)^{2n+1}_{+}={\cal B}_{2n+1}$ will be purely imaginary.  Further
$\partial_P{\cal B}_{2n+1}$ will involve only even powers of $P$ and hence
will be real.  
Thus write now 
\be
{\cal B}_{2n+1}=\sum_0^nb_{nj}P^{2j+1};\,\,\partial_P{\cal B}_{2n+1}=
\sum_0^n(2j+1)b_{nj}P^{2j}
\label{LM}
\ee
and we have
\be
\frac{d}{dT_n}\Im{\cal F}=-\frac{1}{\epsilon}\dot{X}_n=\frac{1}{\epsilon}
\sum_0^n(2j+1)b_{nj}P^{2j}
\label{LN}
\ee
Then the condition $P=iQ$ leads to a compatible KdV situation
(\ref{LN}) and further
\be
\dot{P}_n=\frac{dP}{dT_n}=\partial{\cal B}_n=\sum_0^n\partial (b_{nj})
P^{2j+1}
\label{LO}
\ee
which is realistic (and imaginary).
\\[3mm]\indent 
Now we note that there
is danger here of a situation where $\Re P=0$ implies $\Re S=0$
which in turn would imply $|\psi|^2=1$ (going against the philosophy
of keeping $|\psi|^2$ as a fundamental variable) and this is one reason
we will need $dKdV_{\epsilon}$ with (\ref{OF}) - (\ref{OH}).
Thus in general
\be
S=\sum_1^{\infty}T_n\lambda^n-\sum_1^{\infty}\frac{\partial_n F}{n}\lambda^{-n};
\label{LP}
\ee
$$P=\lambda-\sum_1^{\infty}\frac{F_{1n}}{n}\lambda^{-n};\,\,{\cal B}_m
=\lambda^n-\sum_1^{\infty}\frac{F_{mn}}{n}\lambda^{-n}$$
and for KdV (with $\lambda=ik$) it follows from the residue formula
(cf. \cite{cj}) that
\be
F_{nm}=F_{mn}=Res_P\left(\lambda^md\lambda^n_{+}\right)
\label{LQ}
\ee
that $F_{m,2n}=0$ and from a $\bar{\partial}$ analysis (cf. \cite{ci,cj})
\be
\partial_jF=\frac{j}{2i\pi}\int\int \zeta^{j-1}\bar{\partial}_{\zeta}S
d\zeta\wedge d\bar{\zeta}
\label{LR}
\ee
The $\partial_jF$ and $F_{1j}$ can be computed explicitly as in \cite{cj}
and in particular $F_{1,2n}=0$ with ($P^2-U=-k^2$)
\be
F_{1,2n-1}=(-1)^n\left(\frac{U}{2}\right)^n\prod_1^n\frac{2j-1}{j}
\label{LS}
\ee
A further calculation along the same lines also shows that $F_{2n}=\partial_
{2n}F=0$ for KdV.
Generally $F$ will be real along with the $F_{mn}$ and we recall that the
expression for ${\cal B}_{2m+1}$ arising from (\ref{LP}) is an alternate
way of writing (\ref{LM}).  For $\lambda=ik,\,\,P$ and ${\cal B}_{2m+1}$
will be purely imaginary but $S$ could be complex 
via $\sum_1^{\infty}T_n\lambda^n$ since all powers $\lambda^n
=(ik)^n$ will occur in (\ref{LP}).  Thus $\Re S\not=0$ and we have a
perfectly respectable situation, provided the $T_{2n}$ are real.
However $T_{2n}$ imaginary as in KP1 (cf. \cite{ca}), or 
as in (\ref{B}), would imply 
$\Re S=0$ and $|\psi|^2=1$ which is not desirable.
Another problem is that if $\Re S\not= 0$ is achieved via the times then
$|\psi|^2\sim exp[(1/\epsilon)\sum T_{2n}\lambda^{2n}]$ will not necessarily
be $\leq 1$.  Thus if $dKdV_{\epsilon}$ is not used
this would seem to force
a KP situation with $\partial_P{\cal B}_n$ real, and $P$ complex
(with constraint $\lambda=P+\sum U_{n+1}P^{-n}$ satisfying $\lambda^2$
real); then the equation $\dot{P}_n=\partial{\cal B}_n$ does not require
$\partial{\cal B}_n$ to be real.  However some care with $\lambda$ is indicated
since $P^2-U=\lambda^2\sim-k^2$ would require also $P^2$ to be real if
in fact this equation were used to define $S$ via $S_X=P$ and would force
us back to KdV with $\Re S=0$ and $|\psi|^2=1$.  Indeed $\partial_2\psi
=\lambda^2\psi$ means $L^2_{+}\psi=\lambda^2\psi$ but we know $L^2\psi
=\lambda^2\psi$ in general so this implies $L^2=L^2_{+}$ which is KdV.
Hence we would have to go back to (\ref{A}) 
with KPI and be sure to interpret it as an
eigenvalue equation $i\hbar\partial_{\tau}\psi={\cal H}\psi =E\psi$
(we should also label $\psi=\psi_E$ as in \cite{fa},
with variable $\lambda$ divorced from $E$ entirely).
\\[3mm]\indent
Thus one could temporarily reject dKdV, substitute dKPI, and 
continue with (\ref{LK}) with Hamiltonian type equations
$\dot{P}_n=\partial{\cal B}_n$ and $\dot{X}_n=-\partial_P{\cal B}_n$
for a Hamiltonian $H_n=-{\cal B}_n\,\,(n\geq 2)$.  We could envision
a symplectic form $dX\wedge dP$ on some phase space $M$ built from
variables $(X,P)$ with ${P}$ possibly complex.
In fact there is no admonition now to prevent taking $P$ real except that
$\Im P=-1/\chi$ so $\Im P\not= 0$ is mandatory. 
In order to connect with \cite{fa} and Section 2
we recall in Section 3 one goes from $\partial^2-U$ to $P^2-U$ whereas
in Section 2 we begin with $\epsilon^2\partial^2_X-V$
and pass to $P^2-V$; thus
only $U\sim V$ is required.  The ensuing phase space contains an element 
$P$ not in the original quantum mechanical problem but $P$ is connected
to $S\,\,(P=S_X)$ and to $F$.  Given a connection of $F$ to ${\cal F}$ as in
(\ref{AKQ}) we expect that $M$ could possibly be expressed in terms of
$X$ and ${\cal F}$ or ${\cal F}$ alone.
The symplectic forms of Section 2 involve $\chi=
|\psi|^2$ and $\xi=S/2\epsilon$ for $\psi=|\psi|exp(iS/\epsilon)$, or
$\bar{\psi}$ and $\psi$, or $\chi$ and $\phi=(1/2)exp(-2iS/\epsilon)$,
where a real $S$ is used to measure phase.  In the KP formalism $\psi
=exp(S/\epsilon)=exp(\Re S/\epsilon)exp(i\Im S/\epsilon)$ and we recall
that $P\chi=-1$ in (\ref{UU}) corresponds here to $(\bullet\spadesuit)\,\,
(\Im P)\chi=-1$ as in (\ref{AGQ}).  Also $|\psi|=exp(\Re S/\epsilon)$ with
$\chi=exp(2\Re S/\epsilon)=2\Re{\cal F}=-1/\Im P$ (from (\ref{AGQ})) and
$\xi=\Im S/2\epsilon$ is related to $\phi$ via $2\phi=exp(-4i\xi)$ which
implies $log(2\phi)=-4i\xi$ or $\xi=(i/4)log(2\phi)$ again as in (\ref{VV}).
Then again $\psi^2\phi =(1/2)\chi$ as in (\ref{VV}), leading to $\delta\xi
\wedge \delta\chi\sim(i/2)\delta\phi\wedge\delta(\psi^2)\sim (i/2)\delta
\bar{\psi}\wedge\delta\psi$ as in (\ref{XX}), where $\xi=(1/2\epsilon)\Im
S$ and 
$\chi=|\psi|^2=exp(2\Re S/\epsilon)$ with $\chi=2\Re{\cal F}=-1/\Im P$
and $\Im{\cal F}=-(X/\epsilon)$. 
Thus $X=-\epsilon\Im{\cal F}$ and $\Im P=-(1/2\Re{\cal F})=-(1/\chi)$ 
appear to be fundamental variables and one looks for $P$ in terms of 
$\chi$ and $\phi$ for example, or ${\cal F}$.  
The condition $P=iQ$ for dKdV was seen to be inappropriate as 
above so $\Re P=0$ with $Q=-(1/2\Re{\cal F})$ seems untenable
(but see below for $dKdV_{\epsilon}$ where it does work).
In any event $X=-\epsilon\Im{\cal F}$ 
and
for a realistic situation $\Re S\not= 0$ with $\Im P\not= 0$ suggesting
$\Re P\not= 0$ as well; this would imply that $P$ is genuinely complex
(neither purely real nor imaginary).  Further the condition $\chi\Im P=-1$
from (\ref{AGQ}) is a constraint specifying
\be
\frac{2}{\epsilon}\Re S=log\left(\frac{-1}{\Im P}\right)=\pi i-log(\Im P)
\label{LY}
\ee
If one is
considering $X$ and $P$ as fundamental variables in the HJ approach then
(cf. (\ref{LK})) $\dot{X}_n=-\partial_P{\cal B}_n$ should be real and 
$\dot{P}_n=\partial {\cal B}_n$ will be complex.  From (\ref{LY}) we obtain
\be
\frac{2}{\epsilon}\Re{\cal B}_n=
\frac{2}{\epsilon}\Re\partial_nS=-\frac{\Im\partial_nP}{\Im P}=-\frac
{\Im\partial{\cal B}_n}{\Im P}
\label{LZ}
\ee
The presence of all these
$\epsilon$ terms suggests strongly that we modify
the HJ theory and not drop the $\epsilon$
terms.  We have also been forgetting the origin of (\ref{A}), namely 
${\cal H}\psi_E=E\psi_E$ for example, corresponding to $i\hbar\partial_
{\tau}\psi_E=E\psi_E$ for suitable $\tau$ and this is another constraint.
In any case the variables are severely constrained 
for dKPI and one does not seem to
get nice formulas; hence we will momentarily reject this also.

\subsection{HJ with $dKdV_{\epsilon}$}

Let us now use some of the $dKdV_{\epsilon}$ expressions.
In view of (\ref{OF}) - (\ref{OH}) there is now no problem with $\Re S^0=0$
while happily 
$\Re S^1\not= 0$ and $|\psi|^2\leq 1$ is realistic.  The equation
(\ref{YR}) applies now with variations as in (\ref{one}); we cannot write
$ik\sim\tilde{P}(1+q\tilde{P}^{-2})^{\frac{1}{2}}$ 
however since the $\epsilon P_X$ terms will be missing.
Note here also that $P
=ik-\sum_1^{\infty}P_n(ik)^{-n}$ inverts (\ref{YR}) with $P_n=0$ for
$n$ even ($P_n=F_{1n}/n$ here - cf. \cite{cj} where there is an index
shift in the $P_n$); this shows that $P=iQ$.  
If we write (\ref{two}) for $dKdV_{\epsilon}$ with $\lambda=ik$ and
$\tilde{q}=q+\epsilon\hat{q}$ one obtains
(cf. (\ref{YR}))
\be
ik=
P\left[1+\sum{\frac{1}{2} \choose m}q^mP^{-2m}\right]+
\epsilon\sum_0^{2n+1}\hat{b}_{n(2j+1)}P^{2j+1}+
\label{OI}
\ee
$$
+\epsilon P^1\left[1-
2\sum_1^{\infty}\left(\frac{\frac{1}{2}\cdots
(\frac{3}{2}-m)}{(m-1)!}\right)q^mP^{-2m}
\right]+\epsilon\hat{q}\sum_1^{\infty}m{\frac{1}{2} \choose m}q^{m-1}
P^{-2m+1}+$$
$$+\epsilon\sum_0^{\infty}\left(\frac{2m+1)(m+1)}
{\psi}\right)(\epsilon\partial)^
{-2m-1}\left[\frac{P_X\psi}{P^2}\right]
+O(\epsilon^2)$$
and similarly we can expand (\ref{three}) to get
\be
(ik)^{2n+1}_{+}\sim\tilde{{\cal B}}_{2n+1}={\cal B}_{2n+1}+\epsilon
{\cal B}^1_{2n+1}={\cal B}_{2n+1}(P)+\epsilon\sum_0^{2n+1}\hat{b}_{n(2j+1)}
P^{2j+1}+
\label{OJ}
\ee
$$+\epsilon\left(
P^1\sum_0^{2n+1}(2j+1)b_{n(2j+1)}P^{2j}
+P_X\sum_1^{2n+1}j(2j+1)b_{n(2j+1)}P^{2j-1}\right)$$
(cf. (\ref{LM})).
We recall
from (\ref{OH}) that $\Im P^1=0$ while $P=iQ$ is purely imaginary so 
$\tilde{P}=P+\epsilon P^1=\epsilon P^1+iQ$ and look at $\dot{X}_{2n+1}$ and
$d\tilde{P}/dT_{2n+1}$.  First $d\tilde{P}/dT_{2n+1}=\dot{P}_{2n+1}
+\epsilon dP^1/dT_{2n+1}=\partial\tilde{{\cal B}}_{2n+1}$ is indicated
and one expects $\dot{P}_{2n+1}=\partial{\cal B}_{2n+1}(P)$ to hold.  Hence
we would want 
\be
\frac{dP^1}{dT_{2n+1}}=\partial{\cal B}^1_{2n+1}=
\partial\sum_0^{2n+1}\hat{b}_{n(2j+1)}P^{2j+1}+
\label{ZZ}
\ee
$$+\partial\left[P^1b_{n1}+
\sum_1^{2n+1}(2j+1)b_{n(2j+1)}P^{2j-1}(P^1P+jP_X)\right]$$
Note here that the constraint $|\psi|^2\Im\tilde{P}=-1\equiv |\psi|^2\Im P
=-1$ and this can be written $exp(2S^1)\Im P=-1$.  For $P=iQ$ imaginary and
$S^1,\,P^1$ real we obtain
\be
2S^1+log(\Im P)=i\pi\Rightarrow 2P^1=-\frac{\Im P_X}{\Im P}\Rightarrow
P_X=-2PP^1
\label{OT}
\ee
(applied to the $\hat{V}$ formula after (\ref{OD}) this gives $\hat{V}=0$).
This means that ${\cal B}^1_{2n+1}$ can be simplified to read
\be
{\cal B}^1_{2n+1}=P^1\left[b_{n1}-\sum_1^{2n+1}(2j+1)(2j-1)b_{n(2j+1)}P^{2j}
\right]+\sum_0^{2n+1}\hat{b}_{n(2j+1)}P^{2j+1}
\label{ZY}
\ee
Since $P^1$ is real we must have
\be
\frac{dP^1}{dT_{2n+1}}=\partial\left\{P^1\left[b_{n1}+\sum_1^{2n+1}
(4j^2-1)b_{n(2j+1)}P^{2j}\right]\right\};
\,\,0=\partial\sum_0^{2n+1}\hat{b}_{n(2j+1)}
P^{2j+1}
\label{ZV}
\ee
which seem to place constraints on $\hat{b}_{n(2j+1)}$ and $P$ ($\hat{b}_
{n(2j+1)}=0$ would work which is perhaps analogous to $\hat{V}=0$ above).
Next for
$\dot{X}_{2n+1}$ we consider 
e.g. $\dot{X}_{2n+1}=-\partial\tilde{{\cal B}}_{2n+1}/
\partial\tilde{P}$ for $\tilde{P}=\epsilon P^1+iQ$ so as in complex 
function theory one expects $\partial_z=\partial_x-i\partial_y$ or
$\partial_{\tilde{P}}=\epsilon^{-1}\partial_{P^1}+\partial_P$ here
($\partial_y=i\partial_{iy},\,\,\partial_{P^1}=\epsilon\partial_
{\epsilon P^1}$).  Then
\be
\dot{X}_{2n+1}=-\partial_P{\cal B}_{2n+1}-\partial_{P^1}{\cal B}^1_{2n+1}
-\epsilon\partial_P{\cal B}^1_{2n+1}
\label{ZX}
\ee
which would imply
\be
\dot{X}_{2n+1}=\partial_P{\cal B}_{2n+1}-\partial_{P^1}{\cal B}^1_{2n+1}=
-2b_{n1}-\sum_1^{2n+1}(2j)(2j+1)b_{n(2j+1)}P^{2j}+
\label{ZW}
\ee
$$
+\sum_0^{2n+1}(2j+1)\hat{b}_{n(2j+1)}P^{2j}$$
which is real as desired.  There is also an apparent constraint
$\partial_P{\cal B}^1_{2n+1}=0=-P^1\sum_1^{2n+1}(2j)(4j^2-1)b_{n(2j+1)}P^{2j-1}
+\sum_0^{2n+1}(2j+1)\hat{b}_{n(2j+1)}P^{2j}$
which however should 
probably be combined with a term $\epsilon^2\cdot\epsilon^{-1}
\partial_{P^1}{\cal B}^2_{2n+1}$ in a continued expansion 
(along with $\hat{b}_{n(2j+1)}=0$) and may not
represent a severe constraint.
Alternatively $X\to\tilde{X}=X+\epsilon X^1$ with $X^1$ imaginary and
$dX^1/dT_{2n+1}=-\partial_P{\cal B}^1_{2n+1}$ could be envisioned 
(although with difficulty).
We do not pursue this however since
in fact the HJ theory is not crucial
here as far as $\tilde{P}=P+\epsilon P^1$ is concerned.  Given $S=S^0
+\epsilon S^1,\,\,F=F^0+\epsilon F^1$ (or simply
$F^0$), we know $\tilde{P}=P+\epsilon P^1$
is correct and that is all that is needed for the formulas of \cite{fa} - also
for $V=V^0+\epsilon \hat{V}$ with $\hat{V}=0$ mandated later.  
Thus we take now $\lambda^2=-E$ (cf. Section 4.1) and specify 
$dKdV_{\epsilon}$.  We can
still label $\psi$ as $\psi_E$ but now one imagines a $T_2\sim\tau$ variable 
inserted e.g. via $\psi=\psi(X,T_{2n+1})exp(E\tau/i\hbar)\,\,(n\geq 0)$
with $i\hbar\psi_{\tau}=E\psi$ and $\epsilon^2\psi''-V\psi=-E\psi=\lambda^2
\psi\,\,(V=V^0+\epsilon \hat{V}$ as in (\ref{OD}) and $V=V(X,T_{2n+1})$).

\subsection{Formulas based on Section 2}

Consider ${\cal F}=(1/2)\psi\bar{\psi}+(X/i\epsilon)$ with $\psi=
exp[(1/\epsilon)S^0+S^1],\,\,\Re S^0=0$ as in (\ref{OG}), and 
$|\psi|^2=exp(2\Re S^1)$ as in (\ref{OF}).  Here
\be
S^0=i\left[\sum_1^{\infty}T_{2n+1}(-1)^nk^{2n+1}+
\sum_1^{\infty}S_{2n}(-1)^nk^{-2n+1}\right]
\label{OK}
\ee
and explicitly
\be
{\cal F}=\frac{1}{2}exp\left[\sum_1^{\infty}\left(\frac{F^0_{(2m+1)(2n+1)}}
{(2m+1)(2n+1)}\right)(-1)^{m+n+1}k^{-2(m+n+1)}\right]+\frac{X}{i\epsilon}
\label{OL}
\ee
Thus the $\epsilon$ ``problem" has been removed from the $|\psi|^2$ term but
$\epsilon$ still occurs as a scale factor with $X$.
Look now at (\ref{ADQ}) with $P$ replaced by $\tilde{P}$ to obtain
$|\psi|^2\Im\tilde{P}=-1$ which in view of the $\epsilon$ independence
of $|\psi|^2$ suggests that $\Im P^1=0$ which in fact is true from (\ref{OH}).
Thus $|\psi|^2\Im P=-1$ as before but $P=S^0_X$ now.  Next for $\phi=
\bar{\psi}/2\psi$ we have $\phi=(1/2)exp[-(2i/\epsilon)\Im S]$ and 
$S^0$ is imaginary as in (\ref{OK}) with $S^1$ real as indicated in (\ref{OH}).
Consequently
\be
\phi=\frac{1}{2}exp\left[-\frac{2i}{\epsilon}\Im S^0\right]=
\label{OM}
\ee
$$=\frac{1}{2}exp\left[-\frac{2i}{\epsilon}\left(\sum_0^{\infty} 
(-1)^nT_{2n+1}k^{2n+1}+\sum_1^{\infty}(-1)^nS_{2n}k^{-2n+1}\right)\right]$$
\indent
One can also return to the discussion at the end of Section 4.1 and
suggest again that $X=-\epsilon\Im {\cal F}$ and (for $P=iQ$)
\be
P=i\Im\tilde{P}=i\Im P^0=i\Im P=iQ^0=-\frac{i}{|\psi|^2}=-\frac{i}{2\Re
{\cal F}}
\label{ON}
\ee
are fundamental variables.  Note also from (\ref{OK}), $log(2\phi)=-(2/\epsilon)
S^0$, so
\be
log(2\phi)=-4i\xi=-\frac{2i}{\epsilon}\left(\sum (-1)^nT_{2n+1}k^{2n+1}+
\sum (-1)^nS_{2n}k^{-2n+1}\right)
\label{OOO}
\ee
From $dX\wedge dP$ we obtain now as a possibly fundamental symplectic form
\be
dX\wedge dP=-\epsilon d(\Im{\cal F})\wedge\frac{i}{2(\Re{\cal F})^2}
d(\Re {\cal F})=-\frac{i\epsilon}{2(\Re{\cal F})^2}d(\Im{\cal F})\wedge
d(\Re{\cal F})
\label{OP}
\ee
which has a certain charm and
seems intrinsically
related to the duality idea based on ${\cal F}$
(note this not $dX\wedge d\tilde{P}$, which would involve an additional
term $dX\wedge dP^1$, where a relation to $dX\wedge dP$ could be 
envisioned via $P^1=-(1/2)\partial_Xlog\,P$).  
The constraint $|\psi|^2\Im P=-1$ becomes
\be
exp[2\Re S^1]\Im S^0_X=-1\equiv -1=
\label{OQ}
\ee
$$=exp\left[\sum\left(\frac{F^0_{2m+1)(2n+1)}}{(2m+1)(2n+1)}\right)(-1)^{m+n+1}
k^{-2(m+n+1)}\right]\cdot\left[kX+\sum\partial_XS_{2n}(-1)^nk^{-2n+1}\right]$$
where $\partial_XS_{2n}=-\partial_X(\partial_{2n-1}F/(2n-1))=-
(F_{1,2n-1}/(2n-1)$.  This also seems to be realistic and possibly
interesting.
\\[3mm]\indent
Let us compute the form $\omega =\delta\xi\wedge\delta\chi$ from (\ref{XX}) in
one of its many forms.  First recall $S^0$ is imaginary and $S^1$ is real
with $log(2\phi)=-(2/\epsilon)S^0=-4i\xi$ and $\chi=|\psi|^2=exp(2S^1)$.
Therefore formally, via $\xi=-(i/2\epsilon)S^0$, we have
\be
\omega=\delta\xi\wedge\delta\chi= -\left(\frac{i\chi}{\epsilon}\right)
\delta S^0\wedge\delta S^1
\label{OR}
\ee
The difference here from (\ref{OP}) for example is that the term $X=-\epsilon
{\cal F}$ has no relation to $S^0$ or $S^1$ a priori.  One is tempted to
write e.g. $X=\int (dX/d\psi)d\psi$, based on the strong dependence $X=
X(\psi)$ and $\psi=\psi(X)$ with $1=(dX/d\psi)\psi'$; this seems to lead
to 
\be
X=\int\frac{1}{\psi'}d\psi=\int\frac{\epsilon}{\tilde{P}}\frac{d\psi}{\psi}
=\int\left(\frac{\epsilon}{\tilde{P}}\right)d\,log\psi
\label{OS}
\ee
which would imply perhaps $dX=(\epsilon/\tilde{P})d\,log\psi$ leading to
$dX=(d\tilde{S}/\tilde{P})$.  This would be 
different from $\partial_X\tilde{S}=\tilde{P}$
and might be exploitable.

\section{QUANTUM MECHANICS VIA CLASSICAL THEORY}
\renewcommand{\theequation}{5.\arabic{equation}}\setcounter{equation}{0}

We go now to a fascinating series of papers by Olave (cf. \cite{oa}) which
develop quantum mechanics (QM) via the density matrix and classical
structures.  We will simply look at some equations here and refer to
\cite{oa} for an extensive philosophy.  There are a number of possible
connections with the theory of \cite{ca,fa} which we hope to explore
further in another paper.
One starts from three axioms:
\begin{itemize}
\item
Newtonian mechanics is valid for all particles which constitute the
systems in the ``ensemble".
\item
For an isolated system the joint probability density function is
conserved, i.e. $(\clubsuit)\,\,(d/dt)F(x,p,t)=0$.
\item
The Wigner-Moyal infinitesimal transformation defined by (7.1) below
is adequate for the description of any non-relativistic quantum system 
(note the emphasis here on the infinitesimal aspect).
\end{itemize}
\be
\rho\left(x+\frac{\delta x}{2},x-\frac{\delta x}{2},t\right)=\int F(x,p,t)
exp\left(i\frac{p\delta x}{\hbar}\right)dp
\label{MB}
\ee
(we use small $x$ here to follow \cite{oa}
and for comparison with \cite{ca,fa} or the rest of this paper 
one should convert this
to large $X$).
Using these axioms one produces non-relativistic QM (sort of); in any event
this is the most thorough and penentrating attempt we have seen using the
density matrix.  One can find a few objections at various points but 
heuristically at least the treatment seems very attractive and is possibly
correct (but we leave such judgements for people more versed in physics). 
First from $(\clubsuit)$ one has
\be
\frac{dF}{dt}=\frac{\partial F}{\partial t}+\frac{dx}{dt}\frac{\partial F}
{\partial x}+\frac{dp}{dt}\frac{\partial F}{\partial p}
\label{MC}
\ee
One can then use $dx/dt=p/m$ and $dp/dt=f=-\partial V/\partial x$ in
(\ref{MC}); then multiplying the resulting equation by $exp(ip\delta x/
\hbar)$ and integrating there results
\be
-\frac{\partial \rho}{\partial t}+\frac{i\hbar}{m}\frac{\partial^2\rho}
{\partial x\partial(\delta x)}-\frac{i}{\hbar}\delta V\rho=0
\label{MD}
\ee
where one writes
\be
\frac{\partial V}{\partial x}\delta x=\delta V(x)=V\left(x+
\frac{\delta x}{2}\right)
-V\left(x-\frac{\delta x}{2}\right)
\label{ME}
\ee
and uses the fact that $[F(x,p,t)exp(ip\delta x/\hbar)]_{-\infty}^{\infty}=0$.
Changing variables via $y=x+(\delta x/2)$ and $y'=x-(\delta x/2)$ (\ref{MD}) 
can be rewritten as
\be
\left\{\frac{\hbar^2}{2m}\left[\frac{\partial^2}{\partial y^2}-\frac{\partial^2}
{\partial (y')^2}\right]-[V(y)-V(y')]\right\}\rho(y,y',t)=
-i\hbar\frac{\partial}{\partial t}\rho(y,y',t)
\label{MF}
\ee
which is called Schr\"odinger's first equation for the density function.
\\[3mm]\indent
Now one assumes
\be
(A)\,\,\rho(y,y',t)=\psi^*(y',t)\psi(y,t)\,\,\,\,(B)\,\,\psi(y,t)=R(y,t)
exp\left(\frac{iS(y,t)}{\hbar}\right)
\label{MG}
\ee
where $\psi$ will be called a probability amplitude (perhaps not the best
terminology).  Next one expands $\rho$ in (\ref{MG}A) in terms of
$\delta x$ to get
\be
\rho\left(x+\frac{\delta x}{2},x-\frac{\delta x}{2}\right)=
\label{MH}
\ee
$$=\left\{R(x,t)^2-\left(\frac{\delta x}{2}\right)^2\left[\left(
\frac{\partial R}{\partial x}\right)^2-R(x,t)\frac{\partial^2R}{\partial x^2}
\right]\right\}exp\left(\frac{i}{\hbar}\delta x\frac{\partial S}{\partial x}
\right)$$
Putting this in (\ref{MD}) yields
\be
\left[\frac{\partial(R^2)}{\partial t}+\frac{\partial}{\partial x}
\left(R^2\frac{S_x}{m}\right)\right]+
\label{MI}
\ee
$$+\frac{i\delta x}{\hbar}\left\{\frac{\hbar^2}{2mR^2}\frac{\partial}{\partial
x}\left[\left(\frac{\partial R}{\partial x}\right)^2-R\frac{\partial^2R}
{\partial x^2}\right]+\frac{\partial}{\partial x}\left[\frac{1}{2m}
\left(\frac{\partial S}{\partial x}\right)^2+V+\frac{\partial S}{\partial t}
\right]\right\}=0$$
Equating real and imaginary parts one arrives at
\be
(A)\,\,\frac{\partial \chi}{\partial t}+\frac{\partial}{\partial x}\left(\chi
\frac{S_x}{m}\right)=0\,\,\,\,(B)\,\,\frac{\partial}{\partial x}\left[
\frac{1}{2m}\left(\frac{\partial S}{\partial x}\right)^2+V+\frac{\partial S}
{\partial t}-\frac{\hbar^2}{2mR}\frac{\partial^2R}{\partial x^2}\right]=0
\label{MJ}
\ee
where we have changed notation a bit in using 
$(\spadesuit)\,\,\chi=R^2=lim\,\rho
[x+(\delta x/2),x-(\delta x/2)]$ as $\delta x\to 0$; eventually we will
want to refer to $S_x$ as $P$ etc. with subsequent adjustment to 
$\epsilon$ or $\hbar$ as in Section 4. 
Here $\chi$ is the standard 
probability density in configuration space.  Now rewrite (\ref{MJ}B) as
\be
\frac{1}{2m}\left(\frac{\partial S}{\partial x}\right)^2+V+
\frac{\partial S}{\partial t}-\frac{\hbar^2}{2mR}\frac{\partial^2R}
{\partial x^2}=c=0
\label{ML}
\ee
($c=0$ via consideration of a free particle).  This is then equivalent to
\be
\frac{\hbar^2}{2m}\frac{\partial^2\psi}{\partial x^2}-V(x)\psi=-i\hbar\frac
{\partial\psi}{\partial t}
\label{MMM}
\ee
which will be called Schr\"odinger's second equation (if $\psi=Rexp(iS/\hbar)$
is substituted in (\ref{MMM}) one obtains (\ref{MJ}B)).  
In summary:  If one
can write (\ref{MG}A) then $\psi$ given by (\ref{MG}B) satisfies (\ref{MMM})
along with the equation of continuity (\ref{MJ}A).
\\[3mm]\indent
One then defines operators $(\bullet)\,\,\hat{p}'=-i\hbar\partial/
\partial(\delta x);\,\,\hat{x}'=x$ based on
\be
\bar{p}=\lim_{\delta x\to 0}\left[-i\hbar\frac{\partial}{\partial(\delta x)}
\int F\,exp\left(i\frac{p\delta x}{\hbar}\right)dxdp\right];
\label{MN}
\ee
$$\bar{x}=lim_{\delta x\to 0}\int xF\,exp\left(i\frac{p\delta x}{\hbar}\right)
dxdp$$
(the apostrophe indicates operators acting on the density function; without
the apostrophe operators act on the ``probability amplitude").
\\[3mm]\indent {\bf REMARK 5.1.}$\,\,$ In \cite{oa} one summarizes by stating
that the result of the operation upon the density function $\rho$ of the
momentum and position operators defined by $(\bullet)$ represents,
respectively, the mean values for momentum and position for the ensemble
components.  Remarks are also made to the effect that mean values are
calculated within the limit $\delta x\to 0$ as in (\ref{MN}) so that
calculation of the density function for infinitesimally close points
can be done without loss of generality.  This does not imply that only
the element for which $\delta x=0$ contributes.  The kinematic evolution
of the density function is governed by (\ref{MD}) which mixes all
contributions.  The limits for determining $\chi$ or mean values must
be taken after (\ref{MD}) is solved.  Note also (cf. (\ref{MH}))
\be
\bar{p}=lim_{\delta x\to 0}\left[-i\hbar\frac{\partial}{\partial(\delta x)}
\int Fexp\left(i\frac{p\delta x}{\hbar}\right)dxdp\right]=\int R^2\left(
\frac{\partial S}{\partial x}\right)dx
\label{MO}
\ee
which is behind the terminology of continuity equation for (\ref{MJ}A).  
\\[3mm]\indent
In order to find the momentum operator action on the ``probability
amplitude" one rewrites (\ref{MO}) as
\be
\bar{p}=lim_{\delta x\to 0}\left[-i\hbar\int\frac{\partial}{\partial(\delta x)}
\left\{\psi^*\left(x-\frac{\delta x}{2},t\right)\psi\left(x+\frac{\delta x}{2}
\right)\right\}dx\right]
\label{MP}
\ee
which, after some calculation, leads to
\be
\bar{p}=\Re\left\{\int\psi^*(x,t)\left(-i\hbar\frac{\partial}{\partial x}\right)
\psi(x,t)dx\right\}
\label{MQ}
\ee
The same procedure applies to the position operator and one notes that
the Hermitian character of these operators is automatic.  Thus we have defined
\be
\hat{p}\psi(x,t)=-i\hbar\frac{\partial}{\partial
x}\psi(x,t);\,\,\hat{x}\psi(x,t)=
x\psi(x,t)
\label{MR}
\ee
as usual and the Hamiltonian operator is defined as
\be
\hat{H}\psi(x,t)=\left[\frac{p^2}{2m}+V(x)\right]\psi(x,t)=i\hbar\frac
{\partial}{\partial t}\psi(x,t)
\label{MS}
\ee
Then Schr\"odinger's second equation has the operator form $\hat{H}\psi
=i\hbar(\partial/\partial t)\psi$.  Note that one has now $[\hat{x},\hat{p}]
=i\hbar$ and it can be shown that in fact 
$(\clubsuit\clubsuit)\,\,\Delta x\Delta p\geq \hbar/2$
(Heisenberg uncertainty).  However one has $[\hat{x}',\hat{p}']=0$ so
$\Delta x\Delta p\geq 0$ in this context.  This is natural since no
hypotheses about the mean square deviations associated to the classical
function $F$ were made.  The uncertainty principle $(\clubsuit\clubsuit)$ 
results from writing the density function as the product (\ref{MG}A).
Thus instead of representing a fundamental property of nature the
uncertainty principle simply represents a limitation of the description
based on Schr\"odinger's second equation (\ref{MMM}).  Consequently quantum
mechanics as developed above is only applicable to problems where the 
density function can be decomposed as in (\ref{MG}A).
Further one remarks that the dispersion relations do not impose any
constraint upon the behavior of nature but only upon our capacity
to describe nature by means of quantum theory.  If e.g. $\Delta q
\Delta p<\hbar/2$ in some situation then quantum theory does not apply
or will not give good results.
\\[3mm]\indent {\bf REMARK 5.2.}$\,\,$ 
We note that for a stationary problem $\partial_t\chi=0$ the condition
(\ref{MJ}) becomes $\chi P=c$ which is compatible with (\ref{UU}) where
$\chi P=-1$.  The equation (\ref{ML}) on the other hand becomes
$(\spadesuit\spadesuit)\,\,(P^2/2m)+V-(\hbar^2/2m)(\sqrt{\chi}''/\sqrt
{\chi})=0$ which corresponds to $\dot{\xi}=0$ in (\ref{RR}).
\\[3mm]\indent
Now returning to (\ref{ML}) one can consider this as a 
Hamilton Jacobi (HJ) equation for
one particle subject to an effective potential
\be
V_{eff}=V(x)-\frac{\hbar^2}{2mR}\frac{\partial^2R}{\partial x^2}
\label{MT}
\ee
Thus formally with initial condition $P=S_X$ one can write
\be
\dot{P}=-\partial_xV_{eff}(x)=-\partial_x\left[V-\frac{\hbar^2}{2mR}
\frac{\partial^2R}{\partial x^2}\right]
\label{MU}
\ee
Recall the HJ theory gives, for transformations $H(p,q)\to H'(\hat{P})$,
a HJ equation $S_t+H(q,(\partial S/\partial q),t)=0$.  Here $H(q,
S_q,t)\sim (1/2)P^2+V_{eff}$ so dynamically one has the Hamilton equation
$\dot{p}=-\partial H/\partial q=-\partial_xV_{eff}$
(we use $p\equiv P$).
One thinks of $S=S(q,
\hat{P},t)$ with $\hat{P}$ constant 
and $S$ is called a generating function.  For
conservative systems $S=S(q,\hat{P})-Et$ so one gets $H(z,\partial S/\partial
q)=E=H'(\hat{P})$.  We refer here also to (\ref{RR}) where 
\be
\hbar\dot{\xi}=\frac{\dot{S}}{2}=\frac{\hbar^2}{4m}\frac{\sqrt{\chi}''}
{\sqrt{\chi}}-\frac{V}{2}-\frac{\hbar^2}{m}\frac{P^2}{4\hbar^2}=
-\frac{1}{2}\left(\frac{P^2}{2m}+V_{eff}\right)\Rightarrow
\label{MV}
\ee
$$\Rightarrow \frac{\dot{P}}{2}=-\frac{1}{2}\partial_x\left(\frac{P^2}
{2m}+V_{eff}\right)\Rightarrow\dot{P}=-\partial_xV_{eff}$$
when $P$ and $x$ are considered as independent variables.
This material should be related to the HJ theory sketched earlier for
dKP and/or dKdV.
The integration of (\ref{MU}) with $P=S_x$ will give a series of
trajectories equivalent to the force lines associated to $V_{eff}$.  The
resolution method goes as follows:  First Schr\"odinger's equation
must be solved in order to obtain the ``probability amplitudes"
referring to the ensemble.  Then the effective potential, which will act
as a statistical field for the ensemble is constructed.  This potential
should not be considered as a real potential but a ficticious one which
acts as a field in reproducing through trajectories the statistical
results of the original equation (\ref{ML}).
\\[3mm]\indent {\bf REMARK 5.3.}$\,\,$
The Wigner-Moyal transformation (\ref{MB}) has a formal inverse
\be
F(x,p,t)=\int \rho\left(x+\frac{\delta x}{2},x-\frac{\delta x}{2}\right)
exp\left(-i\frac{p\delta x}{\hbar}\right)d(\delta x)
\label{MW}
\ee
Of course one knows that such a formula does not give a positive
function $F$ and in fact it cannot be used here since $\delta x$ is
an infinitesimal quantity.  The treatment for
ensembles of mixed states can be done via
\be
\rho\left(x+\frac{\delta x}{2},x-\frac{\delta x}{2},t\right)=
\sum_nW_n\psi^*_n\left(x-\frac{\delta x}{2},t\right)\psi_n\left(x+
\frac{\delta x}{2},t\right)
\label{MZZ}
\ee
where the $W_n$ are the statistical weights (we omit details here).
The operator ordering problem can
be thought of in terms of mapping a commutative ring into a noncommutative
ring and this is discussed at length in \cite{oa}.  One obtains an
unambiguous procedure giving the same result as Weyl ordering.
\\[3mm]\indent 
Regarding thermodynamic behavior now,
the idea is to let the systems $(S)$ composing an ensemble interact
with a neighborhood $(O)$ called the heat bath.  The interaction is
considered sufficiently feeble so as to allow one to write a Hamiltonian
$H(q,p)$ for $S$ not depending on the degrees of freedom of $(O)$.
The system $O$ is necessary only as a means of imposing its temperature
$T$ upon $S$.  Now in a state of equilibrium there is a canonical probability
distribution $F(q,p)=Cexp(-2\beta H(q,p))$ where $2\beta=1/K_BT$ with
$K_B$ being the Boltzmann constant, $T$ the absolute temperature, and 
$C$ some normalization constant.  The Hamiltonian may be written
\be
H(q,p)=\sum_1^N\frac{p_n^2}{2m_n}+V(q_1,\cdots,q_N)
\label{ND}
\ee
Using the Wigner-Moyal transformation we get
\be
\rho\left(q+\frac{\delta q}{2},q-\frac{\delta q}{2}\right)=C\int e^{-2\beta
[\sum(p_n^2/2m_n)+V]}\cdot e^{(i\sum p_n\delta q_n/\hbar)}\prod dp_j
\label{NE}
\ee
which yields
\be
\rho=C_1e^{-2\beta V(q)}\cdot e^{-\sum(m_n/4\beta \hbar^2)(\delta q_n)^2}
\label{NF}
\ee
Evidently this ``characteristic function" is a solution to
\be
-\sum_1^N\frac{\hbar^2}{m_n}\frac{\partial^2\rho}{\partial q_n\partial
(\delta q_n)}+\sum_1^N\frac{\partial\rho}{\partial q_n}(\delta q_n)\rho=0
\label{NG}
\ee
Writing $\rho$ in the form (\ref{MG}A) one may take
\be
\psi(q,t)=\sqrt{C_3}e^{-\beta V(q)}\cdot e^{-\frac{iEt}{\hbar}}
\label{NH}
\ee
yielding
\be
\rho=C_3e^{-2\beta[V(q)+(1/8)\sum (\delta q_n)^2(\partial^2 V/\partial q_n^2)]}
\label{NI}
\ee
Comparing with (\ref{NF}) we see that around the point $q$ where the 
function is being evaluated 
it is necessary to have $(\bullet\bullet)\,\,(\partial^2V/
\partial q_n^2)_{\delta q_n=0}=m_n/\beta^2\hbar^2$, if we want
$\psi$ as above.
Thus if $(\partial V/\partial q_n)_{\delta q_n=0}=0$ (a mechanical
equilibrium point when combined with $(\bullet\bullet)$)
then one can take $\rho_{eq}=exp(-2\beta V(q))$
with $\rho=\rho_{eq}(q+(\delta q/2))=\rho_{eq}(q-(\delta q/2))$
(Taylor expansion of $V$).
Thus in the present circumstance the characteristic function can be 
considered as a probability density function
when evaluated at points infinitesimally distant from the systems
mechanical equilibrium situation.  This connection between the factorization
of the characteristic function (which allows the derivation of the
Schr\"odinger equation) and the fact that we are dealing with systems
infinitesimally near the mechanical equilibrium points, provide insight
into the validity of the Bohr postulates as formulated in the early days
of QM.
\\[3mm]\indent
Thus one asks first which Schr\"odinger equation is related with  
(\ref{NH}).  Putting this amplitude into the Schr\"odinger
equation $-\sum_1^N(\hbar^2/2m_n)(\partial^2\psi/\partial q_n^2)+V\psi=E\psi$ 
one obtains
\be
\sum_1^N\frac{\beta\hbar^2}{2m_n}\frac{\partial^2V}{\partial q_n^2}+V
-\sum_1^N\frac{\beta^2\hbar^2}{2m_n}\left[\frac{\partial V}
{\partial q_n}\right]^2=E
\label{NJ}
\ee
from which one obtains $E=V(q^0)+NK_BT$ where $q^0$ represents the 
mechanical equilibrium point and $NK_BT$ represents the energy of the
reservoir $O$.  One can now establish a connection between the microscopic
entities of the quantum formalism and the macroscopic description
given by thermodynamics.  Thus define the free energy $F_G(q)=V(q)$ such
that $(\bullet\diamondsuit\bullet)\,\,F_G=-K_BTlog(\psi^*(q)\psi(q))$
(for $C_3=1$ at least).
Writing entropy as $\check S=K_Blog(\psi^*(q)\psi(q))$ we have 
$F_G=-T\check S$.
\\[3mm]\indent
Suppose now
we can write
\be
f(q,p,p',t)=\phi^*(q,2p-p',t)\phi(q,p',t);
\label{NK}
\ee
$$F(q,p,t)=\int_{-\infty}^{\infty}f(q,p,p',t)dp'=\int_{-\infty}^{\infty}
\phi^*(q,2p-p',t)\phi(q,p',t)dp'$$
Then for $\rho$ defined as in (\ref{MB}) one can use the Fourier convolution
theorem to obtain
\be
\rho\left(q+\frac{\delta q}{2},q-\frac{\delta q}{2},t\right)=
T_F\{\phi^*(q,p,t)\}T_F\{\phi(q,p',t)\}
\label{NL}
\ee
where $T_F$ is the Fourier transform.  Now write
\be
\psi\left(q+\frac{\delta q}{2},t\right)=T_F\{\phi(q,p,t)\}=\int
e^{p\delta q/2\hbar}\phi(q,p,t)dp;
\label{NM}
\ee
$$\psi^*\left(q-\frac{\delta q}{2},t\right)=T_F\{\phi^*(q,p,t)\}=
\int e^{p\delta q/2\hbar}\phi^*(q,p,t)dp$$
and one obtains the factorization (\ref{MG}A), i.e. $\rho(q+(\delta q/2),
q-(\delta q/2),t)=\psi^*(q-(\delta q/2),t)\psi(q+(\delta q/2),t)$.
One notes here that (\ref{NM}) is compatible with the identification
$p=-i\hbar(\partial/\partial q)$ since
\be
\psi\left(q+\frac{\delta q}{2},t\right)=\int e^{\frac{\delta q}{2}\frac
{\partial}{\partial q}}\phi(q,p,t)dp=\int\phi\left(q+\frac{\delta q}{2},
p,t\right)dp
\label{NNN}
\ee
Then in \cite{oa} it is shown how (\ref{NM}) creates a bridge between
the present formalism and the ``old quantum theory" of Bohr-Sommerfeld.
\\[3mm]\indent
In the direction of utilizing this framework we suggest a formula
\be
\tilde{{\cal F}}=\frac{1}{2}\psi\left(x+\frac{\delta x}{2}\right)\bar{\psi}
\left(x-
\frac{\delta x}{2}\right)+\frac{X}{i\epsilon}=
\label{OA}
\ee
$$=\frac{1}{2}\rho\left(x+\frac{\delta x}{2},x-\frac{\delta x}{2}\right)
+\frac{X}{i\epsilon}$$
where all $x$ variables here should be capitalized and $t\sim T_n\,\,(n
\geq 2)$.  We recall that in \cite{fa} the equation (\ref{II})
$|\psi|^2=2{\cal F}-(2X/i\epsilon)$ is interpreted as describing the
space variable as a macroscopic thermodynamic quantity with the 
microscopic information encoded in the prepotential.  Then QM can
be reformulated in terms of (\ref{II}) with the Schr\"odinger equation
replaced by the third order equation (\ref{LL}).  Here $\hbar$ can be
considered as the scale of the statistical system (cf. \cite{bd,fa}).  These
comments from \cite{fa} seem completely adaptable to a connection such
as (\ref{OA}) with the theory of \cite{oa}.
Now suppose we define $\tilde{{\cal F}}$ as in (\ref{OA}) and use
(\ref{MG}B) so that $\chi=R^2=lim\rho[x+(\delta x/2),x-(\delta x/2)]$ as
$\delta x\to 0$, leading to (\ref{MMM}).  Then $\tilde{{\cal F}}\to{\cal F}=
(1/2)\chi+(X/i\epsilon)$ with the mixing equation (\ref{MD}) in the
background.  Further if the $t$ dependence is restricted to $t=t_2$ of
the form $exp(-iEt/\hbar)$ for suitable $t$ then the Schr\"odinger
equation (\ref{MMM}) has the form (\ref{A}).  Now what about 
thermodynamic analogies?  The formula (\ref{NH}) $\psi=
\sqrt{C_3}exp(-\beta V)exp(-iEt/\hbar)$ is attractive for the time
dependence, yielding (\ref{NI}), and eventually then $E=V(q^0)+NK_BT$ where
$q^0$ is a mechanical equilibrium point.  Then for $C_3=1$ one has a free
energy $F_G=-K_BTlog(\bar{\psi}\psi)=-T\hat{S}.$  Note here for $\psi
=Rexp(iS/\hbar)$ one has then $R=exp(-\beta V)$ and $\check S=-Et$ with
$\bar{\psi}\psi=R^2$ so $F_G=-K_BR(2log\,R)=-T\check S$ and $\check S=
2K_Blog\,R$.  In the situation where $\psi=exp(S/\epsilon)$ with complex
$S$ one has $\chi=R^2=|\psi|^2=exp(2\Re S/\epsilon)$ which implies $log\,R^2
=2\Re S/\epsilon=\check S/K_B$ or $\check S=2K_B\Re S/\epsilon$.  However
the ``free energy" in dKP theory is the function $F$ which
stands in a different relation to $S$ than $F_G$ does to $\check S$
(e.g. $S=\sum T_n\lambda^n-\sum (\partial_nF/n)\lambda^{-n}$ as in 
(\ref{four})).
We hope to pick this up again at another time.

\end{document}